\documentclass[aps,prl,twocolumn,groupedaddress,preprintnumbers,showpacs,10pt,amsmath,amssymb]{revtex4}

\usepackage[american]{babel}
\usepackage{epsfig}
\usepackage{longtable}

\newcommand{\qq}[1]{{\lq}#1{\rq}}

\newcommand{\areaof}[1]{\ensuremath{\area_{#1}}}
\newcommand{\critical}[1]{\ensuremath{{#1}^\star}}

\newcommand{\gibbsenergyof}[1]{\ensuremath{\gibbsenergy_{#1}}}
\newcommand{\liquid}[1]{\ensuremath{{#1}_\ell}}

\newcommand{\nuclrateof}[1]{\ensuremath{\nuclrate_{#1}}}
\newcommand{\numberof}[1]{\ensuremath{\absnum_{#1}}}
\newcommand{\onset}[1]{{#1}_\textnormal{on}}
\newcommand{\sat}[1]{\ensuremath{{#1}_\textnormal{s}}}
\newcommand{\stericof}[1]{\ensuremath{\steric_{#1}}}
\newcommand{\surfaceenergyof}[1]{\ensuremath{\surfaceenergy_{#1}}}
\newcommand{\surfacetensionof}[1]{\ensuremath{\surfacetension_{#1}}}

\newcommand{\absnum}{\ensuremath{\mathnormal{N}}}
\newcommand{\Angstrom}{\textnormal{\AA}}
\newcommand{\area}{\ensuremath{\mathnormal{A}}}
\newcommand{\chempot}{\ensuremath{\mathnormal{\mu}}}
\newcommand{\density}{\ensuremath{\mathnormal{\rho}}}
\newcommand{\differential}{\ensuremath{\mathnormal{d}}}
\newcommand{\dropletradius}{\ensuremath{\mathnormal{R}}}
\newcommand{\dropletradiuscap}{\ensuremath{\dropletradius_\textnormal{cap}}}
\newcommand{\Drhovap}{\ensuremath{D^\density_\textnormal{v}}}
\newcommand{\energydeviation}{\ensuremath{\mathnormal{b}}}

\newcommand{\gibbsenergy}{\ensuremath{\mathnormal{G}}}
\newcommand{\hevap}{\ensuremath{\mathnormal{\Delta{}h_{\textnormal{v}}}}}
\newcommand{\inta}{\ensuremath{\mathnormal{n}}}
\newcommand{\intb}{\ensuremath{\mathnormal{i}}}
\newcommand{\isochoric}{\ensuremath{\mathnormal{c_v}}}
\newcommand{\kboltz}{\ensuremath{\mathnormal{k}_\mathnormal{B}}}
\newcommand{\lfkaone}{\ensuremath{\mathnormal{\alpha_1}}}
\newcommand{\lfkatwo}{\ensuremath{\mathnormal{\alpha_2}}}

\newcommand{\LJenergy}{\ensuremath{\mathnormal{\varepsilon}}}
\newcommand{\LJlength}{\ensuremath{\mathnormal{\sigma}}}
\newcommand{\LJmass}{\ensuremath{\mathnormal{m}}}
\newcommand{\lsone}{\textsf{ls\ensuremath{_\textsf{1}}}}
\newcommand{\mass}{\ensuremath{\mathnormal{m}}}
\newcommand{\mult}{\ensuremath{\times}}
\newcommand{\nonaccomodation}{\ensuremath{\mathcal{N}}}
\newcommand{\nuclcoeff}{\ensuremath{\mathnormal{C}}}
\newcommand{\nuclrate}{\ensuremath{\mathnormal{J}}}
\newcommand{\planartension}{\surfacetensionof{\infty}}
\newcommand{\pressure}{\ensuremath{\mathnormal{p}}}
\newcommand{\radius}{\ensuremath{\mathnormal{r}}}
\newcommand{\releasedenergy}{\ensuremath{\mathnormal{q}}}
\newcommand{\steric}{\ensuremath{\mathnormal{s}}}
\newcommand{\supersat}{\ensuremath{\mathcal{S}}}
\newcommand{\surfaceenergy}{\ensuremath{\mathnormal{\zeta}}}
\newcommand{\surfacetension}{\ensuremath{\mathnormal{\gamma}}}
\newcommand{\temperature}{\ensuremath{\mathnormal{T}}}
\newcommand{\Tc}{\ensuremath{\temperature_\textnormal{c}}}
\newcommand{\tolmanlength}{\ensuremath{\mathnormal{\delta}}}
\newcommand{\volume}{\ensuremath{\mathnormal{V}}}
\newcommand{\zeldovich}{\ensuremath{\mathcal{Z}}}

\newcommand{\Jref}{\ensuremath{\sqrt{\LJenergy\LJmass^{-1}}\slash\LJlength^{4}}}

\begin{document}

\preprint{Technical report ITT-M$\Theta$/2008-03/B}

\title{ Modification of the classical nucleation theory based on   
        molecular simulation data for surface tension, critical nucleus
        size, and nucleation rate }

\author{Martin Horsch}
\surname{Horsch}
\author{Jadran Vrabec}
\surname{Vrabec}
\thanks{Author to whom correspondence should be addressed: J.\ Vrabec}
\email{vrabec@itt.uni-stuttgart.de}
\author{Hans Hasse}
\surname{Hasse}
\affiliation{ Institute of Thermodynamics and Thermal Process Engineering,
              Universit{\"a}t Stuttgart, Pfaffenwaldring 9, 70569 Stuttgart, Germany }

\begin{abstract}
Nucleation in supersaturated vapor
is investigated with two series of molecular dynamics simulations in the canonical ensemble. The
applied methods are: \,(a) analysis of critical nuclei
at moderate supersaturations by simulating equilibria of single droplets
with surrounding vapors in small systems;
\,(b) simulation of homogeneous nucleation during condensation with
large systems containing $10^5$ -- 10$^{6}$ particles for calculating the nucleation rate of
vapors at high supersaturations.
For the Lennard-Jones fluid, truncated and shifted at 2.5 times the size parameter, it
is shown that the classical nucleation theory underestimates
both the nucleation rate and the size of the critical
nucleus. A surface property corrected modification of this theory is proposed
to consistently cover data on the surface tension of the curved interface,
the critical nucleus size, and the nucleation rate.
\end{abstract}

\pacs{64.60.qe, 68.03.Cd, 82.60.Nh}

\maketitle

\section{Introduction}

Homogeneous nucleation during condensation of supersaturated vapors is a well-studied topic;
however, it is not yet fully understood despite its general importance.
The most widespread modeling approach is still the classical
nucleation theory (CNT) \cite{Gibbs28, VW26, Farkas27, FRLP66}.
CNT is an acceptable approximation for some simple
fluids but may yield huge deviations compared to experimental data in other cases \cite{Ford04}.
An important source of error is the assumption that the emerging liquid has
the same thermodynamic properties as the bulk liquid phase \cite{MZLV07}.
Condensation processes of practical interest, e.g.\ in atmospheric science, are usually heterogeneous
or ion-induced and have a more complex mechanism of nucleation \cite{Kulmala03}.
However, to adequately describe
such processes a thorough understanding of the homogeneous case is a prerequisite.

Experimental methods for studying homogeneous nucleation face considerable challenges:
experimentally, a homogeneous system without walls or other irregularities
can at best be approximated, a difficulty that is absent in molecular simulation. 
Furthermore, the experimentally accessible range of the nucleation rate $\nuclrate$ is limited to
comparatively slow processes that are
relatively far from the spinodal \cite{Iland04}. Experimental data on the critical
nucleus size have only recently become available \cite{Debenedetti06}.
In molecular simulation, homogeneous nucleation can straightforwardly be studied by a direct approach
where a supersaturated vapor is observed for some time interval, the emerging nuclei are counted, and
their size is evaluated \cite{YM98, TKTN05a}.
Due to the limitations in computational power, accessible system size and time interval are limited. 
Thus the direct approach can currently only be applied to vapors at high supersaturations,
where nucleation occurs within nanoseconds.

For systems at lower supersaturation it is necessary to follow other, more indirect approaches,
e.g.\ by simulating other ensembles or related systems instead of nucleation
in the supersaturated phase itself, which occurs too slowly.
Key quantities determined from such indirect simulations are the size of the critical nucleus
$\critical{\inta}$ and its Gibbs energy of formation $\critical{\Delta\gibbsenergy}$, and
methods based on transition path sampling also permit a study of kinetic aspects \cite{EMB03}.
Both molecular dynamics (MD) simulation with inserted nuclei in nonequilibrium with the surrounding
supersaturated phase \cite{Zhukovitskii95} or based on transition path sampling \cite{TDP06}
and Monte Carlo (MC) simulation \cite{MZLV07, WGK07} were used for such purposes in the past. 

In the present work, both the direct and an indirect simulation approach were applied
to validate two versions of CNT and to develop a new surface property
corrected (SPC) modification of CNT. A simple model fluid was chosen,
where the intermolecular interactions are described by the Lennard-Jones 
(LJ) potential, truncated and shifted at an intermolecular
distance $\radius = 2.5\LJlength$ \cite{AT87}.
The small cutoff radius leads to relatively
fast simulations and avoids long-range corrections that are hard to estimate
for inhomogeneous systems \cite{IGGUBRM07}.
The truncated and shifted LJ potential (LJTS) defines an important and well studied model fluid that
can be used to describe noble gases and methane very accurately.
A considerable amount of thermodynamic data is available for it and in particular,
the dependence of the surface tension on curvature has been
quantified \cite{VKFH06}.

\section{Classical nucleation theory}

To describe homogeneous nucleation during condensation,
a supersaturated vapor in a volume $\volume$
at the temperature $\temperature$ and a pressure $\pressure$ which is larger than the
saturated vapor pressure $\sat{\pressure}$ is considered. The
quotient $\supersat = \pressure\slash\sat{\pressure}$ is called the supersaturation
of the vapor (with respect to pressure). Starting from a homogeneous vapor
with $\supersat > 1$, nanoscopic droplets
begin to form after some induction time as dispersed nuclei of the emerging liquid phase.
They assume a specific size distribution, and
the critical nucleus size $\critical{\inta}$ is the number of particles $\inta$ where the
Gibbs energy of nucleus formation $\Delta\gibbsenergyof{\inta}$ has its maximal
value $\critical{\Delta\gibbsenergy}$ \cite{VW26}.

The size of the critical nucleus and its energy of formation were discussed by
Gibbs \cite{Gibbs28} from a theoretical standpoint. Given that the number of nuclei with
$\inta$ particles is usually determined in CNT by applying a factor of
$\exp(-\Delta\gibbsenergyof{\inta}\slash\kboltz\temperature)$ to the number of monomers,
an internally consistent approach \cite{BK72} leads to the expression
\begin{equation}
   \Delta\gibbsenergyof{\inta} =
      -(\inta - 1)(\chempot - \sat{\chempot}) + \surfaceenergyof{\inta} - \surfaceenergyof{1}.
\label{eqn:DeltaGn}
\end{equation}
Here, $\surfaceenergyof{\inta}$ is the surface free energy of a nucleus with $\inta$ particles;
$\sat{\chempot}$ and $\chempot$ are the chemical potentials of the saturated and
the supersaturated vapor. In expression (\ref{eqn:DeltaGn}) a negative volume contribution
competes with a positive surface contribution. The difference between the chemical potentials
can be determined from an integral over the pressure
along the isotherm of the metastable vapor
\begin{equation}
   \chempot = \sat\chempot + \int_{\sat{\pressure}}^{\pressure} \frac{\differential\pressure}{\density}.
\end{equation}
Volmer and Weber \cite{VW26} approximated the nucleation rate by
\begin{equation}
   \nuclrate = \nuclcoeff \exp(-\critical{\Delta\gibbsenergy}\slash\kboltz\temperature).
\label{eqn:VW}
\end{equation}
The preexponential coefficient is \cite{FRLP66}
\begin{equation}
  \nuclcoeff = \frac{\critical{\area}\pressure\numberof{1}\zeldovich\nonaccomodation}
                    {\volume\sqrt{2\pi\mass\kboltz\temperature}},
\label{eqn:feder}
\end{equation}
where $\critical{\area}$ represents the surface area of a critical nucleus, $\numberof{1}$ the number of monomers constituting the vapor,
\begin{equation}
   \zeldovich = \sqrt{ \frac{-1}{2\pi\kboltz\temperature}
                       \frac{\partial^2\gibbsenergyof{\inta}}
                            {\partial\inta^2} \Big|_{\inta=\critical{\inta}} },
\end{equation}
is the Zel'dovich factor, and $\mass$ is the mass of a particle.
Furthermore, $\nonaccomodation = \energydeviation^2\slash(\energydeviation^2 + \releasedenergy^2)$
is the thermal nonaccommodation factor which is calculated from
\qq{the energy released on addition of a monomer} to a critical nucleus
\qq{above that needed to maintain the existing temperature} \cite{FRLP66}
\begin{equation}
   \releasedenergy = \hevap - \frac{1}{2}\kboltz\temperature
      - \frac{\partial\surfaceenergyof{\inta}}{\partial\inta}
         \Bigg|_{\inta = \critical{\inta}},
\end{equation}
and the kinetic energy variance
\begin{equation}
   \energydeviation^2 = (\isochoric + \kboltz\slash{}2)\kboltz\temperature^2,
\end{equation}
where $\hevap$ is the bulk enthalpy of vaporization and
$\isochoric$ is the isochoric heat capacity of the vapor \cite{FRLP66}.

CNT is based on the capillarity approximation: the density of a nucleus is assumed to be
the bulk saturated liquid density $\liquid{\density}$ and its surface tension
$\surfacetensionof{\inta}$ to be the surface tension of the planar
interface $\planartension$ \cite{Gibbs28, VW26, Farkas27, FRLP66}.
Nuclei are assumed to be spherical, thus for one containing $\inta$
particles the surface area is
$\areaof{\inta} = \left(6\sqrt{\pi}\inta \slash \liquid{\density}\right)^{2\slash{}3}$.

The surface free energy is related to the surface tension by
$\surfacetension = (\partial\surfaceenergy\slash\partial\area)_{\pressure, \temperature}$. 
The capillarity approximation implies $\surfaceenergyof{\inta} = \planartension\areaof{\inta}$.
An analysis of experimental results presented by Fenelonov \textit{et al.}\ \cite{FKK01} seems to suggest
that for some fluids, the surface tension of small nuclei deviates from $\planartension$
only by factors between 0.92 and 1.14.
Their line of argument is based on the so-called \qq{first fundamental
nucleation theorem} \cite{Ford97}
\begin{equation}
   \left(\frac{\partial\ln\nuclrate}{\partial\ln\supersat}\right)_{\temperature}
      = \critical{\inta} + 1,
   \label{eqn:nucthe}
\end{equation}
according to which the size of the critical nucleus is obtained from the supersaturation
dependence of the nucleation rate at constant temperature. This value of $\critical{\inta}$
is then inserted into the Kelvin equation
\begin{equation}
   \ln\supersat = \frac{8\pi\critical{\surfacetension}}{3\kboltz\temperature}
      \left(\frac{3}{4\pi\liquid{\density}}\right)^{2\slash{}3} \frac{1}{\sqrt[3]{\critical{\inta}}},
   \label{eqn:Kelvin}
\end{equation}
a corollary of standard CNT, to obtain the surface tension of the critical nucleus.
However, the Kelvin equation does not take any dependence of
$\surfacetensionof{\inta}$ on $\inta$ into account.
Hence, it is inconsistent to use this equation for quantifying precisely this size dependence.
The nucleation theorem as given in Eq.\ (\ref{eqn:nucthe}) assumes 
$\differential\chempot = \kboltz\temperature\differential\ln\supersat$, which is a bad approximation
at high temperatures, in particular near the spinodal line. Furthermore, it neglects
the dependence of the preexponential coefficient $\nuclcoeff$ from
Eqs.\ (\ref{eqn:VW}) and (\ref{eqn:feder}) on $\critical{\inta}$, although $\nuclcoeff$
is actually proportional to both $\pressure$ and the surface area of the critical nucleus
(cf.\ Schmelzer \cite{Schmelzer01} for valid forms of the nucleation theorem).

From theoretical considerations \cite{Tolman49, BB99} and simulations \cite{VKFH06, MZLV07}
it can be inferred that the surface tension for interfaces
with a high curvature is actually much lower.
An approximation of the size dependence of the surface tension was given
by Tolman \cite{Tolman49}
\begin{equation}
   \surfacetensionof{\inta} = \frac{\planartension}{1 + 2\tolmanlength\slash\dropletradius},
\label{eqn:Tolman}
\end{equation}
where $\tolmanlength$ is called the Tolman length and $\dropletradius$ is the radius of the nucleus.
Laaksonen, Ford, and Kulmala (LFK) \cite{LFK94} also
proposed a size dependent specific surface energy
\begin{equation}
   \surfaceenergyof{\inta} \slash \areaof{\inta}
      = \planartension (1 + \lfkaone\inta^{-1\slash{}3} + \lfkatwo\inta^{-2\slash{}3}),
\label{eqn:LFK}
\end{equation}
where $\lfkaone$ and $\lfkatwo$ are determined from thermal properties.
This modification leads to predictions which were found to agree better with simulation data
of Tanaka \textit{et al.}\ \cite{TKTN05a} than standard CNT.
In the LFK model, the curvature effect is covered by the single parameter $\lfkaone$, since $\Delta\gibbsenergyof{\inta}$ does not depend on $\lfkatwo$.

\section{Indirect approach: critical nuclei from MD simulations of equilibria}

Phase coexistence methods are an established approach for obtaining equilibrium data from
molecular simulation \cite{MWHC94}.
In the present indirect simulation approach, a single nucleus in equilibrium with a
supersaturated vapor was studied in the canonical ensemble. 
For such simulations it is crucial to choose the relation of the number of particles in the
nucleus to the total number of particles in the system appropriately.
The liquid fraction must be relatively large so that changes in nucleus size
significantly affect the density of the surrounding vapor and the nucleus cannot
evaporate completely because the vapor density increases.
Eventually, an equilibrium is established, where the nucleus contains $\inta$ particles
while the vapor reaches a supersaturated pressure $\pressure > \sat{\pressure}$.

Farkas \cite{Farkas27} pointed out that for a system with $\absnum$ particles
composed of a nucleus containing $\inta$ and a supersaturated vapor containing
$\absnum-\inta$ particles, an equilibrium between the nucleus and the supersaturated
vapor corresponds to the condition $\inta = \critical{\inta}(\pressure, \temperature)$.
This is due to the fact that by definition, the Gibbs energy of nucleus formation
is maximal for $\critical{\inta}(\pressure, \temperature)$ and thus
\begin{equation}
\left(\frac{\partial\gibbsenergyof{\inta}}{\partial\inta}\right)_{\absnum\pressure\temperature} \Bigg|_{\inta=\critical{\inta}(\pressure, \temperature)} = 0,
\end{equation}
holds, which implies that for $\inta = \critical{\inta}(\pressure, \temperature)$, growth and decay
are equally probable.
Since in a nucleation process $\Delta\gibbsenergyof{\inta}$ has a
single maximum \cite{VW26}, this equilibrium condition uniquely identifies
the size of the critical nucleus. 
By minimizing the Helmholtz energy of the system in
an $\absnum\volume\temperature$ simulation, an equilibrium
that characterizes the maximum of its Gibbs energy is established.
The critical nucleus model of Reguera-Reiss nucleation
theory \cite{RR04} reproduces these considerations in an explicit form.

As suggested by Lovett \cite{Lovett07}, the fact that a critical nucleus \qq{can only be
in (stable) equilibrium with a supersaturated vapour in a system with a
finite (small) volume} makes these small systems
where \qq{the thermodynamic analysis is straightforward and the
configurations are easily simulated} an attractive topic for molecular simulation.
Such an approach leads to more accurate data on the critical nucleus,
e.g.\ its size $\critical{\inta}$ or surface tension $\critical{\surfacetension}$,
than the usual method of observing growth and decay of nuclei in nonequilibrium
simulations \cite{Zhukovitskii95}, because it permits straightforward
sampling over a large number of time steps. 
It is also computationally efficient since only small systems are considered.
The molecular simulation of such equilibria is not a novelty in
itself \cite{SSVBR01, VKFH06, SNV07}, but no implications for
critical nuclei were drawn from these studies in the past.
However, Talanquer \cite{Talanquer07} used a similar approach based on
density functional theory for calculating the free energy of formation and the interfacial
density profile of critical nuclei.

\begin{figure}[t]
\includegraphics[width=8.6cm]{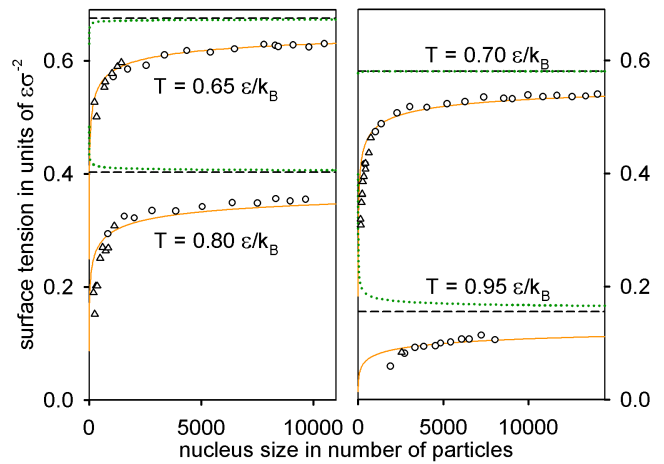}
\caption{
   \label{betI} \label{betV} \label{gimelI}
      (Color)
      Surface tension of the LJTS fluid
      over nucleus size from indirect simulations ($\triangle$ this work, $\circ$ 
      from previous work \cite{VKFH06}) and following 
      standard CNT (dashed lines), 
      the LFK modification of CNT (dots),
      and the new SPC modification of CNT (solid lines).
      The LFK model, which depends on the supersaturated vapor pressure, was evaluated at 
      $\supersat = 2.86$, 2.28, 1.62, and 1.17 for $\temperature = 0.65$, 0.70, 0.80,
      and 0.95 $\LJenergy\slash\kboltz$, respectively.
}
\end{figure}

Simulations in the canonical ensemble based on these considerations can
contribute to the study of nucleation processes indirectly, by reproducing
vapor-liquid equilibria instead of the condensation itself. Such indirect simulations
were conducted for small systems (total number of particles $\absnum < 2\mult10^{4}$)
and properties of the critical nucleus at moderate supersaturations were obtained,
complementing data from previous work \cite{VKFH06}.
Nuclei with $10^2 < \inta < 10^4$ particles were inserted into saturated
or moderately supersaturated vapor phases.
The nucleus size was tracked by applying a version of the cluster criterion of
Rein ten Wolde and Frenkel \cite{WF98} where a particle is considered as belonging
to the nucleus if it has at least four neighbors within a radius of $\radius \leq 1.5\LJlength$.

\begin{figure}[h]
\includegraphics[width=8.6cm]{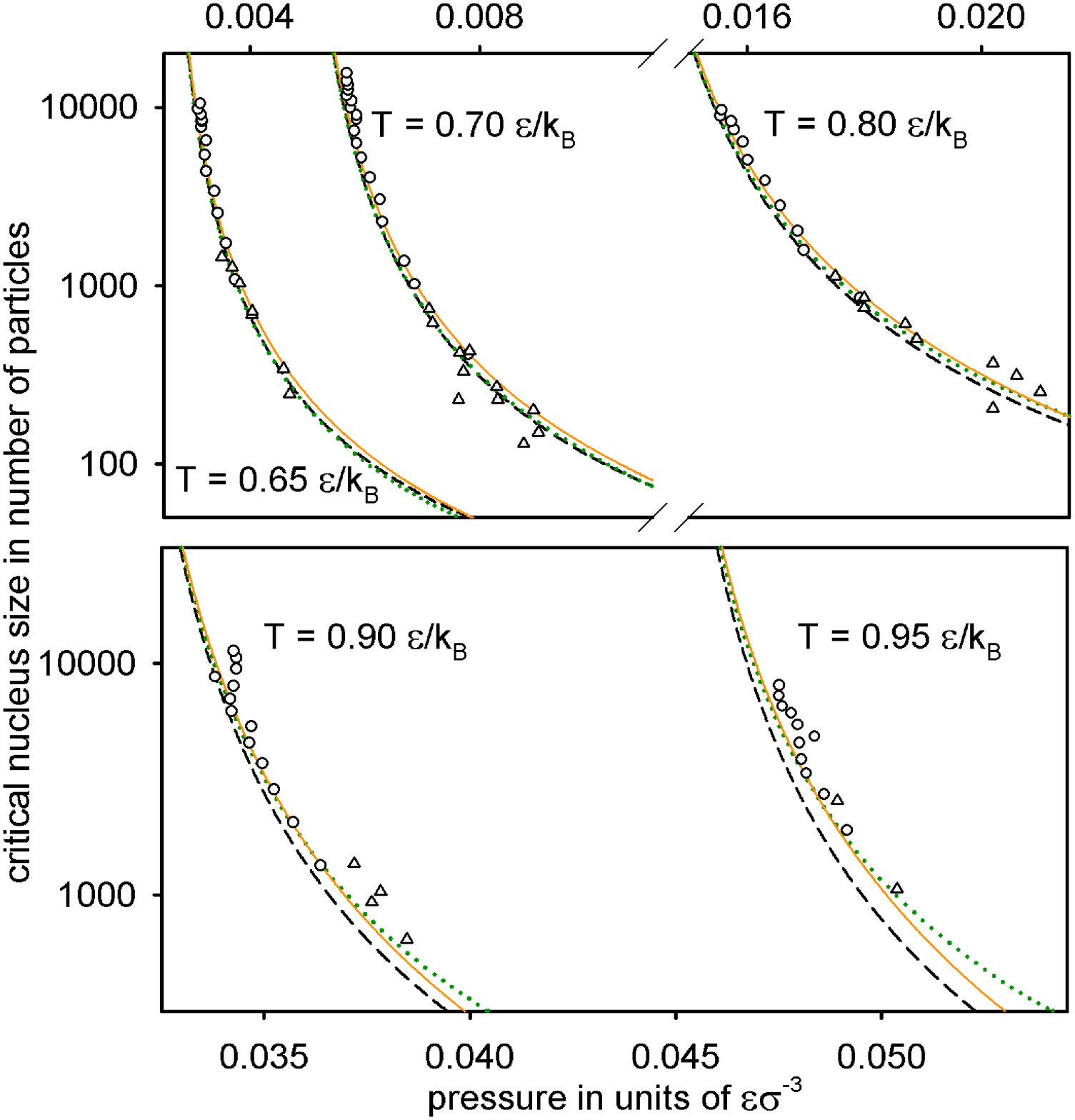}
\caption{
   \label{betII} \label{betIII} \label{betIV} \label{gimelII}
   (Color)
   Critical nucleus size of the LJTS fluid over supersaturated pressure from indirect
   simulations ($\triangle$ this work, $\circ$ 
      from previous work \cite{VKFH06})
   and following standard CNT (dashed lines),
   the LFK modification of CNT (dots), 
   and the new SPC modification of CNT (solid lines).
}
\end{figure}

Surface tension and size of the critical nucleus were determined for six temperature
values between 0.65 and 0.95 $\LJenergy\slash\kboltz$,
cf.\ Fig.\ \ref{gimelI} as well as Tabs.\ \ref{tab:nstar1} and \ref{tab:nstar2}.
As usual, all numerical results are given in terms of the mass $\LJmass$ of a single particle
and the two potential parameters $\LJlength$ (size) and $\LJenergy$ (energy).
The planar interface surface tension $\planartension$ of the LJTS fluid,
given by \cite{VKFH06}
\begin{equation}
   \frac{\planartension\LJlength^{2}}{\LJenergy} = 2.08
      \left(1 - \frac{\temperature}{\Tc}\right)^{1.21},
\end{equation}
with $\Tc = 1.0779$ $\LJenergy\slash\kboltz$,
is represented by horizontal lines in Fig.\ \ref{gimelI} because standard CNT assumes
the surface tension of the curved interface to be size independent and therefore
equal to $\planartension$.
The surface tension, calculated from the normal component of
the Irving-Kirkwood pressure tensor \cite{IK50}, is significantly reduced for small
nuclei when compared to $\planartension$.
Standard CNT neglects this effect, and as shown in Fig.\ \ref{gimelI},
the LFK expression for $\surfaceenergyof{\inta}$ given by Eq.\ (\ref{eqn:LFK}) may even lead to values of
$\surfacetensionof{\inta}$ for small nuclei which are unphysical,
i.e. $\surfacetensionof{\inta} > \planartension$, and increase for smaller nuclei. 

\begin{table}[b]
\caption{
   \label{tab:nstar1}
   Size of the critical nucleus (in number of particles) and
   its surface tension (in units of $\LJenergy\slash\LJlength^2$)
   at low temperatures (given in units of $\LJenergy\slash\kboltz$)
   from simulation in comparison to theories -- bold values
   are taken from earlier work \cite{VKFH06}
}
\begin{ruledtabular}
\begin{tabular}{ll||rr|rrr}
   $\temperature$ &
   $\supersat$ & $\critical{\inta}$
               & $\critical{\surfacetension}$
               & $\critical{\inta}(\textnormal{CNT})$
               & $\critical{\inta}(\textnormal{LFK})$ 
               & $\critical{\inta}(\textnormal{SPC})$ 
   \\ \hline 
   \textbf{0.65}  &
   \textbf{1.226}
         & \textbf{10500} & \textbf{0.630}
         &  7800 &  7800 &  8500 \\ \textbf{0.65} &
   \textbf{1.337} &  \textbf{3400} & \textbf{0.610}
         &  2700 &  2700 &  3000 \\ \textbf{0.65} &
   \textbf{1.420} &  \textbf{1700} & \textbf{0.585}
         &  1600 &  1500 &  1700 \\ 0.65 &
   1.461 &  1300 & 0.590
         &  1200 &  1200 &  1400 \\ 0.65 &
   1.512 &  1000 & 0.578 
         &   950 &   940 &  1100 \\ 0.65 &
   1.594 &   690 & 0.553 
         &   670 &   660 &   750 \\ 0.65 &
   1.599 &   720 & 0.563
         &   650 &   640 &   730 \\ 0.65 &
   1.813 &   340 & 0.501 
         &   320 &   320 &   360 \\ 0.65 &
   1.856 &   250 & 0.527 
         &   290 &   280 &   320 \\ \hline
   \textbf{0.70}  &
   \textbf{1.179} & \textbf{15600} & \textbf{0.535}
         &  8900 &  8700 &  9600 \\ \textbf{0.70} &
   \textbf{1.306} &  \textbf{2300} & \textbf{0.507}
         &  2100 &  2100 &  2300 \\ \textbf{0.70} &
   \textbf{1.420} &  \textbf{1000} & \textbf{0.474}
         &   930 &   930 &  1100 \\ 0.70 &
   1.474 &   740 & 0.463
         &   690 &   690 &   790 \\ 0.70 &
   1.586 &   420 & 0.418
         &   410 &   420 &   470 \\ 0.70 &
   1.621 &   430 & 0.408
         &   360 &   370 &   410 \\ 0.70 &
   1.722 &   230 & 0.363
         &   260 &   260 &   290 \\ 0.70 &
   1.816 &   130 & 0.319
         &   200 &   190 &   220 \\ 0.70 &
   1.869 &   150 & 0.309
         &   170 &   170 &   190 \\ \hline
   \textbf{0.80}  &
   \textbf{1.125} & \textbf{9700} & \textbf{0.355}
         & 8000 & 8300 & 9100 \\ \textbf{0.80} &
   \textbf{1.179} & \textbf{3900} & \textbf{0.334}
         & 2900 & 3100 & 3400 \\ \textbf{0.80} &
   \textbf{1.227} & \textbf{1600} & \textbf{0.325}
         & 1500 & 1600 & 1800 \\ 0.80 &
   1.264 & 1100 & 0.308
         & 1000 & 1100 & 1200 \\ 0.80 &
   1.301 &  750 & 0.264
         &  730 &  790 &  870 \\ 0.80 &
   1.352 &  610 & 0.270
         &  490 &  530 &  580 \\ 0.80 &
   1.366 &  500 & 0.251
         &  450 &  490 &  520 \\ 0.80 &
   1.460 &  200 & 0.190
         &  260 &  280 &  290 \\ 0.80 &
   1.518 &  250 & 0.152
         &  190 &  220 &  220 
\end{tabular}
\end{ruledtabular}
\end{table}

\begin{table}[t]
\caption{
   \label{tab:nstar2}
   Size of the critical nucleus (in number of particles) and
   its surface tension (in units of $\LJenergy\slash\LJlength^2$)
   at high temperatures (given in units of $\LJenergy\slash\kboltz$)
   from simulation in comparison to theories -- bold values
   are taken from earlier work \cite{VKFH06}
}
\begin{ruledtabular}
\begin{tabular}{ll||rr|rrr}
   $\temperature$ &
   $\supersat$ & $\critical{\inta}$
               & $\critical{\surfacetension}$
               & $\critical{\inta}(\textnormal{CNT})$
               & $\critical{\inta}(\textnormal{LFK})$ 
               & $\critical{\inta}(\textnormal{SPC})$ 
   \\ \hline 
   \textbf{0.85}  &
   \textbf{1.102} &  \textbf{8800} & \textbf{0.268}
         &  7100 &  7700 &  8400 \\ \textbf{0.85} &
   \textbf{1.136} &  \textbf{3000} & \textbf{0.250}
         &  3200 &  3500 &  3800 \\ 0.85 &
   1.152 &  2200 & 0.222
         &  2400 &  2600 &  2800 \\ \textbf{0.85} &
   \textbf{1.168} &  \textbf{2200} & \textbf{0.240}
         &  1800
         &  2000
         &  2200 \\ 0.85 &
   1.227 &   840 & 0.202
         &   800 &   920 &  970 \\ 0.85 &
   1.253 &  1300 & 0.224
         &   600 &   700 &   730 \\ 0.85 &
   1.264 &   680 & 0.157
         &   540
         &   630
         &   650 \\ 0.85 &
   1.340 &   450 & 0.127
         &   290 &   340 &   340 \\ 0.85 &
   1.424 &   250 & 0.072
         &   170 &   210 &   190 \\ \hline
   \textbf{0.90}  &
   \textbf{1.086} &  \textbf{7000} & \textbf{0.191}
         &  5600 &  6400 &  6800 \\ \textbf{0.90} &
   \textbf{1.111} &  \textbf{3700} & \textbf{0.173}
         &  2700 &  3200 &  3400 \\ \textbf{0.90} &
   \textbf{1.134} &  \textbf{2100} & \textbf{0.165}
         &  1600 &  2000 &  2000 \\ 0.90 &
   1.182 &  1400 & 0.132
         &   700 &   890 &   880 \\ 0.90 &
   1.198 &   930 & 0.118
         &   570 &   730 &   700 \\ 0.90 &
   1.201 &  1000 & 0.139
         &   550 &   700 &   680 \\ 0.90 &
   1.223 &   650 & 0.033
         &   410 &   540 &   511 \\ \hline
   \textbf{0.95}  &
   \textbf{1.067} & \textbf{7200} & \textbf{0.114}
         & 4400 & 5600 & 5700 \\ \textbf{0.95} &
   \textbf{1.074} & \textbf{6100} & \textbf{0.107}
         & 3300 & 4300 & 4300 \\ \textbf{0.95} &
   \textbf{1.077} & \textbf{5400} & \textbf{0.102}
         & 3000 & 3900 & 3900 \\ \textbf{0.95} &
   \textbf{1.082} & \textbf{3400} & \textbf{0.092}
         & 2400 & 3300 & 3200 \\ \textbf{0.95} &
   \textbf{1.086} & \textbf{4800} & \textbf{0.100}
         & 2200 & 2900 & 2900 \\ 0.95 &
   1.099 & 2600 & 0.084
         & 1400 & 2000 & 1900 \\ 0.95 &
   1.104 & 1900 & 0.059
         & 1300 & 1800 & 1700 
\end{tabular}
\end{ruledtabular}
\end{table}

Simulation results from our group
for $\critical{\inta}$ are compared in Fig.\ \ref{gimelII} to theoretical
values. The LFK modification is in better agreement with simulation data over
the entire studied temperature range than standard CNT
which consistently underestimates $\critical{\inta}$ and leads
to particularly large deviations at high temperatures. 

\begin{figure}[b]
\includegraphics[width=8.6cm]{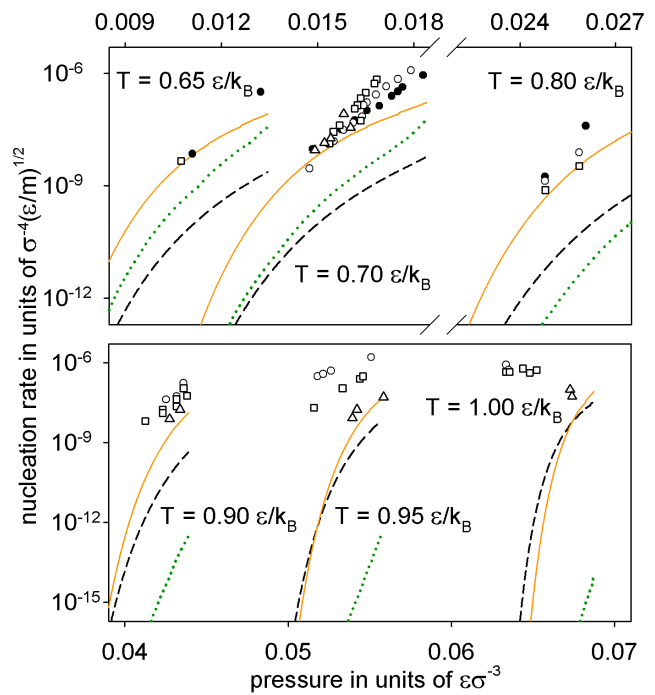}
\caption{
   \label{betVI} \label{betVII} \label{betVIII} \label{gimelIII}
   (Color)
   Nucleation rate of the LJTS fluid over supersaturated pressure
   from the present direct simulations 
   for different threshold values ($\bullet$ $\intb = 25$, $\circ$ $\intb = 50$,
   $\square$ $\intb \in \{75, 100\}$, $\triangle$ $\intb \geq 150$) and following
   standard CNT (dashed lines),
   the LFK modification of CNT (dots),
   and the new SPC modification of CNT (solid lines).
}
\end{figure}

\section{Direct approach: MD simulation of homogeneous nucleation}

A series of direct simulations of the nucleation process was conducted in the
canonical ensemble using the program \lsone{} \cite{BV05}. 
The system size was relatively large ($10^5 < \absnum < 10^6$) and a hybrid cluster criterion
was used to detect the nuclei. 
This criterion combines geometric and energetic approaches with a connectivity analysis
based on graph theory \cite{GRVE07}.
The nucleation rate was determined by defining a threshold $\intb$ and counting the
number $\nuclrateof{\intb}$ of nuclei containing at least $\intb$ particles that emerge
per volume and time \cite{YM98}. According to this method, seven nucleation rate isotherms
were obtained for temperatures between 0.65 and 1 $\LJenergy\slash\kboltz$.


Nucleation rates $\nuclrateof{\intb}$ for different threshold values $\intb$
are compared to theoretical predictions in Fig.\ \ref{gimelIII} as well as Tabs.\ \ref{tab:J1} and \ref{tab:J2}.
The values of $\nuclrateof{\intb}$ are only valid approximations of the actual nucleation
rate if they remain roughly constant for increasing $\intb$ \cite{YM98}.
This is the case for all temperatures except 0.95 and 1 $\LJenergy\slash\kboltz$
where $\critical{\inta}$ is probably larger than all of the chosen threshold values.
With $\critical{\inta} \gg \intb$, the rate of formation for nuclei with $\intb$ or more
particles does not correspond to the nucleation rate, but rather to the velocity at which
the metastable equilibrium which precedes nucleation is established.
For instance, at 0.95 $\LJenergy\slash\kboltz$ and a supersaturation of 1.226,
standard CNT predicts a critical nucleus with 173 particles (LFK: $\critical{\inta} = 293$);
thus, the value of $\nuclrateof{100} = 3 \mult 10^{-7} \sqrt{\LJenergy\mass^{-1}}\slash\LJlength^4$
obtained under these conditions, cf.\ Tab. \ref{tab:J2}, does not describe nucleation but
equilibration. Some other results describe the transition between both regimes, such as
$\nuclrateof{400} = 1 \mult 10^{-7} \sqrt{\LJenergy\mass^{-1}}\slash\LJlength^4$ at
$\temperature = 1$ $\LJenergy\slash\kboltz$ and $\supersat = 1.106$, where the critical size
according to CNT is 328 (LFK: $\critical{\inta} = 755$).

In those cases where $\nuclrateof{\intb}$ clearly represents the actual nucleation process,
consistent deviations were found for standard CNT
which underestimates $\nuclrate$ by two orders of magnitude in all cases.
Such an accuracy should be interpreted as a confirmation of standard CNT
for the LJTS fluid. The LFK modification is in better agreement with simulation data at low
temperatures but leads to larger deviations of $\nuclrate$ at high temperatures.

\begin{table}[t]
\caption{
   \label{tab:J1}
   Nucleation rate (in units of $\Jref$)
   at low temperatures (given in units of $\LJenergy\slash\kboltz$)
   from the present simulations in comparison to theories
}
\begin{ruledtabular}
\begin{tabular}{ll||lr|lll}
   $\temperature$ &
   $\supersat$    & $\nuclrateof{\intb}$ & $\intb$
                  & $\nuclrate(\textnormal{CNT})$ 
                  & $\nuclrate(\textnormal{LFK})$ 
                  & $\nuclrate(\textnormal{SPC})$ \\ \hline
   0.65  &
   4.14  & 5 $\mult$ 10$^{-9}$ & 100 
         & 2 $\mult$ 10$^{-11}$& 2 $\mult$ 10$^{-10}$& 2 $\mult$ 10$^{-9}$ \\ 0.65 &
   4.27  & 7 $\mult$ 10$^{-9}$ &  25 
         & 4 $\mult$ 10$^{-11}$& 4 $\mult$ 10$^{-10}$& 4 $\mult$ 10$^{-9}$ \\ 0.65 &
   5.09  & 3 $\mult$ 10$^{-7}$ &  25
         & 9 $\mult$ 10$^{-10}$& 1 $\mult$ 10$^{-8}$ & 5 $\mult$ 10$^{-8}$ \\ \hline
   0.70  &
   3.06  & 3 $\mult$ 10$^{-9}$ &  50
         & 5 $\mult$ 10$^{-11}$& 1 $\mult$ 10$^{-10}$& 4 $\mult$ 10$^{-9}$ \\ 0.70 &
   3.15  & 1 $\mult$ 10$^{-8}$ & 150
         & 1 $\mult$ 10$^{-10}$& 4 $\mult$ 10$^{-10}$& 8 $\mult$ 10$^{-9}$ \\ 0.70 &
   3.19  & 1 $\mult$ 10$^{-8}$ &  75
         & 1 $\mult$ 10$^{-10}$& 5 $\mult$ 10$^{-10}$& 1 $\mult$ 10$^{-8}$ \\ 0.70 &
   3.22  & 2 $\mult$ 10$^{-8}$ &  50
         & 2 $\mult$ 10$^{-10}$& 6 $\mult$ 10$^{-10}$& 1 $\mult$ 10$^{-8}$ \\ 0.70 &
   3.28  & 8 $\mult$ 10$^{-8}$ & 150
         & 3 $\mult$ 10$^{-10}$& 1 $\mult$ 10$^{-9}$ & 2 $\mult$ 10$^{-8}$ \\ 0.70 &
   3.35  & 1 $\mult$ 10$^{-7}$ &  75
         & 4 $\mult$ 10$^{-10}$& 2 $\mult$ 10$^{-9}$ & 3 $\mult$ 10$^{-8}$ \\ 0.70 &
   3.39  & 2 $\mult$ 10$^{-7}$ &  75
         & 5 $\mult$ 10$^{-10}$& 2 $\mult$ 10$^{-9}$ & 3 $\mult$ 10$^{-8}$ \\ 0.70 &
   3.42  & 3 $\mult$ 10$^{-7}$ &  75
         & 6 $\mult$ 10$^{-10}$& 3 $\mult$ 10$^{-9}$ & 4 $\mult$ 10$^{-8}$ \\ 0.70 &
   3.49  & 3 $\mult$ 10$^{-7}$ &  50
         & 1 $\mult$ 10$^{-9}$ & 4 $\mult$ 10$^{-9}$ & 5 $\mult$ 10$^{-8}$ \\ 0.70 &
   3.55  & 4 $\mult$ 10$^{-7}$ &  50
         & 1 $\mult$ 10$^{-9}$ & 7 $\mult$ 10$^{-9}$ & 6 $\mult$ 10$^{-8}$ 
   \\ \hline
   0.80  &
   1.791 & 2 $\mult$ 10$^{-9}$ & 25
         & 5 $\mult$ 10$^{-12}$& 2 $\mult$ 10$^{-13}$& 5 $\mult$ 10$^{-10}$ \\ 0.80 &
   1.792 & 1 $\mult$ 10$^{-9}$ & 50
         & 6 $\mult$ 10$^{-12}$& 3 $\mult$ 10$^{-13}$& 5 $\mult$ 10$^{-10}$ \\ 0.80 &
   1.792 & 8 $\mult$ 10$^{-10}$& 75
         & 6 $\mult$ 10$^{-12}$& 3 $\mult$ 10$^{-13}$& 5 $\mult$ 10$^{-10}$ \\ 0.80 &
   1.869 & 8 $\mult$ 10$^{-9}$ & 50
         & 5 $\mult$ 10$^{-11}$& 4 $\mult$ 10$^{-12}$& 3 $\mult$ 10$^{-9}$ \\ 0.80 &
   1.869 & 3 $\mult$ 10$^{-9}$ & 75
         & 5 $\mult$ 10$^{-11}$& 4 $\mult$ 10$^{-12}$& 3 $\mult$ 10$^{-9}$ \\ 0.80 &
   1.885 & 4 $\mult$ 10$^{-8}$ & 25
         & 7 $\mult$ 10$^{-11}$& 7 $\mult$ 10$^{-12}$& 5 $\mult$ 10$^{-9}$  \\ \hline
   0.85  &
   1.539 & 3 $\mult$ 10$^{-8}$ & 600
         & 2 $\mult$ 10$^{-11}$& 9 $\mult$ 10$^{-14}$ 
         & 1 $\mult$ 10$^{-9}$ \\ 0.85 &
   1.550 & 2 $\mult$ 10$^{-9}$ & 300
         & 3 $\mult$ 10$^{-11}$& 2 $\mult$ 10$^{-13}$
         & 2 $\mult$ 10$^{-9}$ \\ 0.85 &
   1.560 & 4 $\mult$ 10$^{-9}$ & 600
         & 6 $\mult$ 10$^{-11}$& 3 $\mult$ 10$^{-13}$
         & 2 $\mult$ 10$^{-9}$ \\ 0.85 &
   1.560 & 7 $\mult$ 10$^{-8}$ & 600
         & 6 $\mult$ 10$^{-11}$& 3 $\mult$ 10$^{-13}$
         & 2 $\mult$ 10$^{-9}$ \\ 0.85 &
   1.566 & 1 $\mult$ 10$^{-8}$ & 300
         & 7 $\mult$ 10$^{-11}$& 4 $\mult$ 10$^{-13}$& 3 $\mult$ 10$^{-9}$ \\ 0.85 &
   1.596 & 1 $\mult$ 10$^{-8}$ & 100
         & 2 $\mult$ 10$^{-10}$& 3 $\mult$ 10$^{-12}$
         & 8 $\mult$ 10$^{-9}$ \\ 0.85 &
   1.647 & 2 $\mult$ 10$^{-7}$ & 300
         & 9 $\mult$ 10$^{-10}$& 3 $\mult$ 10$^{-11}$
         & 3 $\mult$ 10$^{-8}$ \\ 0.85 &
   1.656 & 1 $\mult$ 10$^{-7}$ & 100
         & 1 $\mult$ 10$^{-9}$ & 4 $\mult$ 10$^{-11}$& 3 $\mult$ 10$^{-8}$
\end{tabular}
\end{ruledtabular}
\end{table}

\begin{table}[h]
\caption{
   \label{tab:J2}
   Nucleation rate (in units of $\Jref$)
   at high temperatures (given in units of $\LJenergy\slash\kboltz$)
   from the present simulations in comparison to theories
}
\begin{ruledtabular}
\begin{tabular}{ll||lr|lll}
   $\temperature$ &
   $\supersat$    & $\nuclrateof{\intb}$ & $\intb$
                  & $\nuclrate(\textnormal{CNT})$ 
                  & $\nuclrate(\textnormal{LFK})$ 
                  & $\nuclrate(\textnormal{SPC})$ \\ \hline
   0.90  &
   1.31  & 6 $\mult$ 10$^{-9}$ &  75
         & 3 $\mult$ 10$^{-12}$& 9 $\mult$ 10$^{-17}$& 4 $\mult$ 10$^{-11}$ \\ 0.90 &
   1.34  & 1 $\mult$ 10$^{-8}$ & 100
         & 3 $\mult$ 10$^{-11}$& 3 $\mult$ 10$^{-15}$& 6 $\mult$ 10$^{-10}$ \\ 0.90 &
   1.35  & 4 $\mult$ 10$^{-8}$ &  50
         & 7 $\mult$ 10$^{-11}$& 8 $\mult$ 10$^{-15}$& 1 $\mult$ 10$^{-9}$ \\ 0.90 &
   1.36  & 8 $\mult$ 10$^{-9}$ & 200
         & 1 $\mult$ 10$^{-10}$& 2 $\mult$ 10$^{-14}$& 2 $\mult$ 10$^{-9}$ \\ 0.90 &
   1.37  & 5 $\mult$ 10$^{-8}$ &  50
         & 2 $\mult$ 10$^{-10}$& 6 $\mult$ 10$^{-14}$& 4 $\mult$ 10$^{-9}$ \\ 0.90 &
   1.38  & 2 $\mult$ 10$^{-8}$ & 200
         & 4 $\mult$ 10$^{-10}$& 2 $\mult$ 10$^{-13}$& 7 $\mult$ 10$^{-9}$ \\ 0.90 &
   1.39  & 2 $\mult$ 10$^{-7}$ &  50
         & 6 $\mult$ 10$^{-10}$& 3 $\mult$ 10$^{-13}$& 1 $\mult$ 10$^{-8}$ \\ 0.90 &
   1.39  & 1 $\mult$ 10$^{-7}$ &  75
         & 6 $\mult$ 10$^{-10}$& 3 $\mult$ 10$^{-13}$& 1 $\mult$ 10$^{-8}$ \\ 0.90 &
   1.39  & 6 $\mult$ 10$^{-8}$ & 100 
         & 6 $\mult$ 10$^{-10}$& 3 $\mult$ 10$^{-13}$& 1 $\mult$ 10$^{-8}$ \\ \hline
   0.95  &
   1.159 & 2 $\mult$ 10$^{-8}$ & 100
         & 1 $\mult$ 10$^{-13}$& 3 $\mult$ 10$^{-22}$& 2 $\mult$ 10$^{-13}$\\ 0.95 &
   1.172 & 4 $\mult$ 10$^{-7}$ &  50
         & 2 $\mult$ 10$^{-12}$& 3 $\mult$ 10$^{-20}$& 5 $\mult$ 10$^{-12}$\\ 0.95 &
   1.198 & 1 $\mult$ 10$^{-7}$ & 100
         & 7 $\mult$ 10$^{-11}$& 3 $\mult$ 10$^{-17}$& 4 $\mult$ 10$^{-10}$\\ 0.95 &
   1.211 & 8 $\mult$ 10$^{-9}$ & 700
         & 3 $\mult$ 10$^{-10}$& 3 $\mult$ 10$^{-16}$& 2 $\mult$ 10$^{-9}$ \\ 0.95 &
   1.218 & 2 $\mult$ 10$^{-8}$ & 300
         & 5 $\mult$ 10$^{-10}$& 1 $\mult$ 10$^{-15}$& 4 $\mult$ 10$^{-9}$ \\ 0.95 &
   1.221 & 2 $\mult$ 10$^{-7}$ & 100
         & 7 $\mult$ 10$^{-10}$& 2 $\mult$ 10$^{-15}$& 5 $\mult$ 10$^{-9}$ \\ 0.95 &
   1.226 & 3 $\mult$ 10$^{-7}$ & 100
         & 1 $\mult$ 10$^{-9}$ & 4 $\mult$ 10$^{-15}$& 8 $\mult$ 10$^{-9}$ \\ 0.95 &
   1.237 & 2 $\mult$ 10$^{-6}$ &  50
         & 2 $\mult$ 10$^{-9}$ & 2 $\mult$ 10$^{-14}$& 2 $\mult$ 10$^{-8}$ \\ \hline
   1.00  &
   1.039 & 5 $\mult$ 10$^{-8}$ & 150
         & 2 $\mult$ 10$^{-31}$& 3 $\mult$ 10$^{-59}$& 1 $\mult$ 10$^{-39}$\\ 1.00 &
   1.044 & 5 $\mult$ 10$^{-7}$ &  50
         & 4 $\mult$ 10$^{-26}$& 1 $\mult$ 10$^{-50}$& 5 $\mult$ 10$^{-32}$\\ 1.00 &
   1.044 & 4 $\mult$ 10$^{-8}$ & 150
         & 4 $\mult$ 10$^{-26}$& 1 $\mult$ 10$^{-50}$& 5 $\mult$ 10$^{-32}$\\ 1.00 &
   1.057 & 1 $\mult$ 10$^{-7}$ & 150
         & 4 $\mult$ 10$^{-18}$& 9 $\mult$ 10$^{-37}$& 9 $\mult$ 10$^{-21}$\\ 1.00 &
   1.061 & 8 $\mult$ 10$^{-8}$ & 150
         & 1 $\mult$ 10$^{-16}$& 5 $\mult$ 10$^{-34}$& 1 $\mult$ 10$^{-18}$\\ 1.00 &
   1.065 & 4 $\mult$ 10$^{-7}$ &  75
         & 2 $\mult$ 10$^{-15}$& 1 $\mult$ 10$^{-31}$& 7 $\mult$ 10$^{-17}$\\ 1.00 &
   1.072 & 5 $\mult$ 10$^{-7}$ &  75
         & 1 $\mult$ 10$^{-13}$& 3 $\mult$ 10$^{-28}$& 2 $\mult$ 10$^{-14}$\\ 1.00 &
   1.106 & 1 $\mult$ 10$^{-7}$ & 400
         & 1 $\mult$ 10$^{-9}$ & 5 $\mult$ 10$^{-19}$& 4 $\mult$ 10$^{-9}$ \\ 1.00 &
   1.108 & 5 $\mult$ 10$^{-8}$ & 800
         & 2 $\mult$ 10$^{-9}$ & 1 $\mult$ 10$^{-18}$& 6 $\mult$ 10$^{-9}$ 
\end{tabular}
\end{ruledtabular}
\end{table}

\section{Surface property corrected modification of CNT}

As the preceding sections show, standard CNT only predicts the nucleation rate of the
LJTS fluid with an acceptable accuracy, leading to deviations for $\critical{\inta}$;
the LFK modification provides excellent predictions for the critical nucleus size
but not for the temperature dependence of $\nuclrate$. Both theories assume
an inappropriate curvature dependence of the surface tension, although for
this essential property of inhomogeneous systems
a qualitatively correct expression is known since the 1940s \cite{Tolman49}.
With the collected
simulation data on $\critical{\surfacetension}$, $\critical{\inta}$, and $\nuclrate$
over a broad range of temperatures, enough quantitative information is available
to formulate a more adequate modification of CNT.

\begin{figure}[b]
\includegraphics[width=8.6cm]{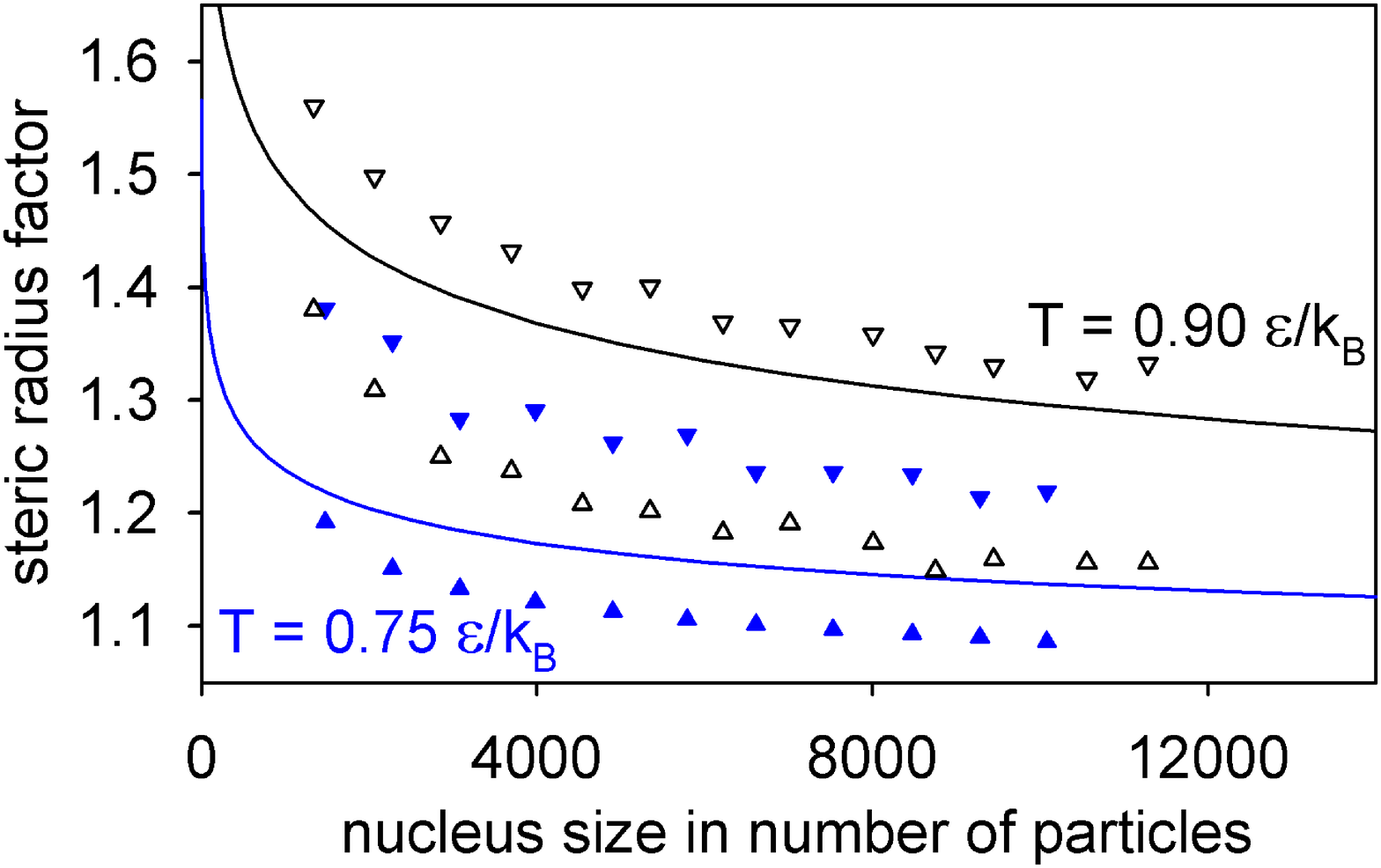}
\caption{
   \label{gimelIV}
   (Color)
   Steric radius factor $\sqrt{\stericof{\inta}}$ in dependence of the nucleus size $\inta$
   at temperatures of 0.75 and 0.90 $\LJenergy\slash\kboltz$ (solid lines);
   values from previous work \cite{VKFH06} are shown for 1 + $\tolmanlength\slash\dropletradiuscap$
   (triangle up) and 1 + $\Drhovap\slash\dropletradiuscap$ (triangle down) at 0.75 (filled symbols) and 0.90
   (empty symbols) $\LJenergy\slash\kboltz$
}
\end{figure}

To correlate the simulation results for $\critical{\surfacetension}$, Eq.\ (\ref{eqn:Tolman})
as proposed by Tolman \cite{Tolman49} was chosen. The quotient
$\tolmanlength\slash\dropletradius$ was assumed to scale with
$\inta^{-1\slash{}3}$, and a fit to the data shown in Fig.\ \ref{gimelI} as well as
Tabs.\ \ref{tab:nstar1} and \ref{tab:nstar2} yields
\begin{eqnarray}
   \tolmanlength\slash\dropletradius
      = \left(\frac{0.7}{1 - \temperature\slash\Tc} - 0.9\right)
         \big\slash \inta^{1\slash{}3}.
\label{eqn:tolmannumber}
\end{eqnarray}
It can be seen in Fig.\ \ref{gimelI} that the two-parameter fit
given by Eqs.\ (\ref{eqn:Tolman}) and (\ref{eqn:tolmannumber}) is sufficient to reproduce
both temperature and size dependence of the surface tension. 

The new SPC modification of CNT is based on a size dependent term for the surface tension,
given by Eqs.\ (\ref{eqn:Tolman}) and (\ref{eqn:tolmannumber}) for the LJTS fluid.
From theoretical considerations \cite{FRLP66} and from simulation \cite{MWB05} it is further known
that the number of particles $\inta$ in the nucleus is insufficient as a reaction coordinate for
nucleation. In particular, the external shape and -- in case of liquid-solid nucleation -- the
internal structure of the nuclei must be taken into account \cite{MWB05, SVFD07}. An MD based analysis of
the emerging crystals in a supercooled LJ liquid by Trudu \textit{et al.}\ \cite{TDP06} showed that for liquid-solid
nucleation, the nuclei are significantly anisotropic: the longest axis of the ellipsoids 
used to approximate the crystal surface was found to be about 1.5 times longer than the shortest axis.

The nonsphericity of the nuclei in a supersaturated vapor, due to fluctuations of the phase boundary,
is represented by a size dependent steric coefficient $\stericof{\inta}$ with
\begin{equation}
   \areaof{\inta} =  \stericof{\inta} \left(6\sqrt{\pi}\inta \slash \liquid{\density}\right)^{2\slash{}3},
\label{eqn:sterical}
\end{equation}
in the present SPC modification of CNT.
The temperature and size dependence of $\stericof{\inta}$
was accounted for by a two-parameter fit
\begin{equation}
   \stericof{\inta} =
      \frac{0.85 \, (1 - \temperature\slash\Tc)^{-1} + (\inta\slash{}75)^{1\slash{}3}
         }{1 + (\inta\slash{}75)^{1\slash{}3}},
\label{eqn:sn}
\end{equation}
adjusted to all simulation results for $\critical{\inta}$ and $\nuclrate$.
This corresponds to an effective increase of the radius according to capillarity
theory $\dropletradiuscap = \sqrt[3]{3 / (4\pi\liquid{\density})}$ by a factor
of $\sqrt{\stericof{\inta}}$. As Fig.\ \ref{gimelIV}
shows, this increase is similar in magnitude to phyiscal properties that express the size of
the phase boundary, such as the Tolman length $\tolmanlength$ and the interface thickness on
the vapor side $\Drhovap$ determined from the average density profile of the nucleus.
Both properties were studied for the LJTS fluid in previous work \cite{VKFH06}.

For all temperatures above 0.162 $\LJenergy\slash\kboltz$, which is far below the triple point
temperature (about 0.61 $\LJenergy\slash\kboltz$), Eq.\ (\ref{eqn:sn}) leads to $\stericof{\inta} > 1$.
For $\temperature\to\Tc$, both the steric coefficient and the thickness of the vapor-liquid interface diverge.
It should be noted that in the case of fluids that cannot
be accurately modeled by the truncated and shifted LJ potential,
the surface tension of small nuclei and the thickness of the phase boundary may require
a different set of parameters for Eqs. (\ref{eqn:tolmannumber}) and (\ref{eqn:sn}). 

As can be seen from Figs.\ \ref{gimelI} -- \ref{gimelIII},
the SPC modification consistently covers the LJTS simulation results
for $\critical{\surfacetension},$ $\critical{\inta},$ and $\nuclrate$.
At the transition between nucleation and spinodal decomposition, the nucleation rates from
simulation exceed the prediction by about an order of magnitude, which may be due to the
particular energy landscape of such processes \cite{BCB07}.
The apparent difference between the values of $\nuclrateof{\intb}$ and all theories
at temperatures of 0.95 and 1 $\LJenergy\slash\kboltz$ is due to the fact that
as discussed above, almost all of these values
describe the velocity of equilibration instead of nucleation.

\section{Application to argon}

Fluid argon can be represented accurately by an LJTS molecular model,
the corresponding potential parameters were
determined as $\LJlength = 3.3916$ \Angstrom{} and $\LJenergy\slash\kboltz = 137.90$ K in
previous work \cite{VKFH06}. Conveniently, experimental nucleation data are available, however,
usually at very low temperatures below the triple point.
The onset pressure $\onset{\pressure}$, defined
as the pressure where the nucleation rate exceeds a certain minimal value $\onset{\nuclrate}$,
was determined for the homogeneous nucleation of argon by Pierce \textit{et al.}\ \cite{PSM71},
Zahoranski \textit{et al.}\ \cite{ZHS95, ZHS99}, and Iland \textit{et al.}\ \cite{IWSK07}.
The onset nucleation rate $\onset{\nuclrate}$ depends on the experimental setup and
was provided (or estimated) by the authors of the respective studies, except for
Pierce \textit{et al.}\ \cite{PSM71}, where we assumed $\onset{\nuclrate} = 10^{22}$ m$^{-3}$s$^{-1}$
as given by Iland \cite{Iland04} for the nucleation rate detected by supersonic nozzles.

\begin{table}[h]
\caption{
   \label{tab:ponset}
   Nucleation onset pressure $\onset{\pressure}$ (in units of kPa) for argon at low
   temperatures (in units of K) from experimental data in comparison to the pressure
   where the assumed onset nucleation rate $\onset{\nuclrate}$ (in units of m$^{-3}$s$^{-1}$) is reached
   according to theories; the data of Pierce \textit{et al.}\ \cite{PSM71} were published in
   graphical form only
}
\begin{ruledtabular}
\begin{tabular}{llr|rrr|l}
  ref.\ & $\temperature$ & $\onset{\pressure}(\textnormal{exp})$
        & $\onset{\pressure}(\textnormal{CNT})$ & $\onset{\pressure}(\textnormal{LFK})$
        & $\onset{\pressure}(\textnormal{SPC})$ & $\onset{\nuclrate}$ \\ \hline
\cite{ZHS95}      &
   48.2 &   0.31
        &   1.2\,\,\,       &   0.83
        &   1.3\,\,\,       &  10$^{6}$ \\
\cite{IWSK07}     &
   48.2 &   1.3
        &   2.1\,\,\,       &   1.4\,\,\,
        &   2.1\,\,\,       &  10$^{13}$ \\
\cite{PSM71}      & 
   55   &  19\,\,\,\,\,\,\,
        &  16\,\,\,\,\,\,\, &   14\,\,\,\,\,\,\,
        &  14\,\,\,\,\,\,\, &  10$^{22}$ \\
\cite{ZHS95}      & 
   55.8 &   0.99
        &   4.8\,\,\,       &   6.0\,\,\,
        &   4.8\,\,\,       &  10$^{6}$ \\
\cite{ZHS99}      & 
   55.9 &   5.28
        &   6.7\,\,\,       &   7.2\,\,\,
        &   6.5\,\,\,       &  10$^{12}$ \\
\cite{IWSK07}     & 
   55.9 &   6.2\,\,\,
        &   7.1\,\,\,       &   7.5\,\,\,
        &   6.9\,\,\,       &  10$^{13}$ \\
\cite{ZHS99}      & 
   60.2 &  11.1\,\,\,
        &  12\,\,\,\,\,\,\, &  13\,\,\,\,\,\,\,
        &  12\,\,\,\,\,\,\, &  10$^{12}$ \\
\cite{ZHS95}      & 
   60.3 &   2.27
        &   9.5\,\,\,       &  11\,\,\,\,\,\,\,
        &   9.3\,\,\,       &  10$^{6}$ \\
\cite{ZHS99}      & 
   62.7 &  12.7\,\,\,
        &  17\,\,\,\,\,\,\, &  17\,\,\,\,\,\,\,
        &  17\,\,\,\,\,\,\, &  10$^{12}$ \\
\cite{PSM71}      & 
   63   &  52\,\,\,\,\,\,\,
        &  34\,\,\,\,\,\,\, &  29\,\,\,\,\,\,\,
        &  30\,\,\,\,\,\,\, &  10$^{22}$ \\
\cite{ZHS99}      & 
   69.9 &  23.9\,\,\,
        &  42\,\,\,\,\,\,\, &  40\,\,\,\,\,\,\,
        &  40\,\,\,\,\,\,\, &  10$^{12}$ \\
\cite{ZHS95}      & 
   85.1 & 114\,\,\,\,\,\,\,
        & 180\,\,\,\,\,\,\, & 180\,\,\,\,\,\,\,
        & 180\,\,\,\,\,\,\, & 10$^{6}$ \\
\cite{PSM71}      & 
   98   & 690\,\,\,\,\,\,\,
        & 570\,\,\,\,\,\,\, & 570\,\,\,\,\,\,\,
        & 540\,\,\,\,\,\,\, & 10$^{22}$
\end{tabular}
\end{ruledtabular}
\end{table}

The comparison between theory and experiment is inconclusive and appears contradictory,
cf.\ Tab.\ \ref{tab:ponset}. Results obtained by Pierce \textit{et al.}\ \cite{PSM71} tend
to confirm CNT. Zahoranski \textit{et al.}\ \cite{ZHS95} observed a nucleation onset at much lower pressures
than all theories. A second study by Zahoranski \textit{et al.}\ \cite{ZHS99} agrees best with the SPC modification,
whereas the correlation proposed by Iland \textit{et al.}\ \cite{IWSK07} confirms LFK.
However, it should be pointed out that at such very low temperatures (60 to 70\% of
the triple point temperature) desublimation processes take place. Therefore, such
severe extrapolations to other parts of the phase diagram are questionable.

\section{Conclusion}

The present simulation results show for the LJTS fluid that
standard CNT underpredicts the nucleation rate $\nuclrate$ at high
supersaturations by about two orders of magnitude. The critical nucleus
size $\critical{\inta}$ is also lower by up to a factor of two at moderate supersaturations. 
The LFK modification is an improvement with respect to $\critical{\inta}$,
but leads to large deviations for $\nuclrate$ at high temperatures. 
A surface property corrected modification of CNT
was presented that takes into account the lower surface tension of
small nuclei and their nonsphericity, effects ignored by standard CNT.
This modification consistently reproduces simulation data for
$\critical{\surfacetension},$ $\critical{\inta},$ and $\nuclrate$ over a wide range of states.


\begin{acknowledgments}
We thank Martin Bernreuther, Karlheinz Schaber, Nicolas Schmidt, Jonathan Walter,
and Andrea Wix for fruitful discussions and Deutsche Forschungsgemeinschaft for
funding Sonderforschungsbereich 716.
The simulations were performed on the HP XC6000 super computer at the Steinbuch
Centre for Computing, Karlsruhe under the grant MMSTP.
\end{acknowledgments}


\begin{thebibliography}{37}
\expandafter\ifx\csname natexlab\endcsname\relax\def\natexlab#1{#1}\fi
\expandafter\ifx\csname bibnamefont\endcsname\relax
  \def\bibnamefont#1{#1}\fi
\expandafter\ifx\csname bibfnamefont\endcsname\relax
  \def\bibfnamefont#1{#1}\fi
\expandafter\ifx\csname citenamefont\endcsname\relax
  \def\citenamefont#1{#1}\fi
\expandafter\ifx\csname url\endcsname\relax
  \def\url#1{\texttt{#1}}\fi
\expandafter\ifx\csname urlprefix\endcsname\relax\def\urlprefix{URL }\fi
\providecommand{\bibinfo}[2]{#2}
\providecommand{\eprint}[2][]{\url{#2}}

\bibitem[{\citenamefont{Gibbs}(1878)}]{Gibbs28}
\bibinfo{author}{\bibfnamefont{J.~W.} \bibnamefont{Gibbs}},
  \emph{\bibinfo{title}{Collected Works. Vol 1. Thermodynamics}}
  (\bibinfo{publisher}{Longmans, Green \& Co.}, \bibinfo{address}{London},
  \bibinfo{year}{1878}).

\bibitem[{\citenamefont{Volmer and Weber}(1926)}]{VW26}
\bibinfo{author}{\bibfnamefont{M.}~\bibnamefont{Volmer}} \bibnamefont{and}
  \bibinfo{author}{\bibfnamefont{A.}~\bibnamefont{Weber}}, \bibinfo{journal}{Z.
  phys. Chem. (Leipzig)} \textbf{\bibinfo{volume}{119}}, \bibinfo{pages}{277}
  (\bibinfo{year}{1926}).

\bibitem[{\citenamefont{Farkas}(1927)}]{Farkas27}
\bibinfo{author}{\bibfnamefont{L.}~\bibnamefont{Farkas}}, \bibinfo{journal}{Z.
  phys. Chem. (Leipzig)} \textbf{\bibinfo{volume}{125}}, \bibinfo{pages}{236}
  (\bibinfo{year}{1927}).

\bibitem[{\citenamefont{Feder et~al.}(1966)\citenamefont{Feder, Russell, Lothe,
  and Pound}}]{FRLP66}
\bibinfo{author}{\bibfnamefont{J.}~\bibnamefont{Feder}},
  \bibinfo{author}{\bibfnamefont{K.~C.} \bibnamefont{Russell}},
  \bibinfo{author}{\bibfnamefont{J.}~\bibnamefont{Lothe}}, \bibnamefont{and}
  \bibinfo{author}{\bibfnamefont{G.~M.} \bibnamefont{Pound}},
  \bibinfo{journal}{Adv. Phys.} \textbf{\bibinfo{volume}{15}},
  \bibinfo{pages}{111} (\bibinfo{year}{1966}).

\bibitem[{\citenamefont{Ford}(2004)}]{Ford04}
\bibinfo{author}{\bibfnamefont{I.~J.} \bibnamefont{Ford}},
  \bibinfo{journal}{Proc. Inst. Mech. Eng. C: J. Mech. Eng. Sci.}
  \textbf{\bibinfo{volume}{218}}, \bibinfo{pages}{883} (\bibinfo{year}{2004}).

\bibitem[{\citenamefont{Merikanto et~al.}(2007)\citenamefont{Merikanto,
  Zapadinsky, Lauri, and Vehkam{\"a}ki}}]{MZLV07}
\bibinfo{author}{\bibfnamefont{J.}~\bibnamefont{Merikanto}},
  \bibinfo{author}{\bibfnamefont{E.}~\bibnamefont{Zapadinsky}},
  \bibinfo{author}{\bibfnamefont{H.}~\bibnamefont{Vehkam{\"a}ki}}, \bibnamefont{and}
  \bibinfo{author}{\bibfnamefont{A.}~\bibnamefont{Lauri}},
  \bibinfo{journal}{Phys. Rev. Lett.} \textbf{\bibinfo{volume}{98}},
  \bibinfo{pages}{145702} (\bibinfo{year}{2007}).

\bibitem[{\citenamefont{Kulmala}(2003)}]{Kulmala03}
\bibinfo{author}{\bibfnamefont{M.}~\bibnamefont{Kulmala}},
  \bibinfo{journal}{Science} \textbf{\bibinfo{volume}{302}},
  \bibinfo{pages}{1000} (\bibinfo{year}{2003}).

\bibitem[{\citenamefont{Iland}(2004)}]{Iland04}
\bibinfo{author}{\bibfnamefont{K.}~\bibnamefont{Iland}}, Ph.D. thesis,
  \bibinfo{school}{University of Cologne} (\bibinfo{year}{2004}).

\bibitem[{\citenamefont{Debenedetti}(2006)}]{Debenedetti06}
\bibinfo{author}{\bibfnamefont{P.~G.} \bibnamefont{Debenedetti}},
  \bibinfo{journal}{Nature} \textbf{\bibinfo{volume}{441}},
  \bibinfo{pages}{168} (\bibinfo{year}{2006}).

\bibitem[{\citenamefont{Yasuoka and Matsumoto}(1998)}]{YM98}
\bibinfo{author}{\bibfnamefont{K.}~\bibnamefont{Yasuoka}} \bibnamefont{and}
  \bibinfo{author}{\bibfnamefont{M.}~\bibnamefont{Matsumoto}},
  \bibinfo{journal}{J. Chem. Phys.} \textbf{\bibinfo{volume}{109}},
  \bibinfo{pages}{8451} (\bibinfo{year}{1998}).

\bibitem[{\citenamefont{Tanaka et~al.}(2005)\citenamefont{Tanaka, Kawamura,
  Tanaka, and Nakazawa}}]{TKTN05a}
\bibinfo{author}{\bibfnamefont{K.~K.} \bibnamefont{Tanaka}},
  \bibinfo{author}{\bibfnamefont{K.}~\bibnamefont{Kawamura}},
  \bibinfo{author}{\bibfnamefont{H.}~\bibnamefont{Tanaka}}, \bibnamefont{and}
  \bibinfo{author}{\bibfnamefont{K.}~\bibnamefont{Nakazawa}},
  \bibinfo{journal}{J. Chem. Phys.} \textbf{\bibinfo{volume}{122}},
  \bibinfo{pages}{184514} (\bibinfo{year}{2005}).

\bibitem[{\citenamefont{van Erp et~al.}(2003)\citenamefont{van Erp, Moroni,
  and Bolhuis}}]{EMB03}
\bibinfo{author}{\bibfnamefont{T.~S.} \bibnamefont{van Erp}},
  \bibinfo{author}{\bibfnamefont{D.} \bibnamefont{Moroni}}, \bibnamefont{and}
  \bibinfo{author}{\bibfnamefont{P.~G.} \bibnamefont{Bolhuis}},
  \bibinfo{journal}{J. Chem. Phys.} \textbf{\bibinfo{volume}{118}},
  \bibinfo{pages}{7762} (\bibinfo{year}{2003}).

\bibitem[{\citenamefont{Zhukovitskii}(1995)}]{Zhukovitskii95}
\bibinfo{author}{\bibfnamefont{D.~I.} \bibnamefont{Zhukovitskii}},
  \bibinfo{journal}{J. Chem. Phys.} \textbf{\bibinfo{volume}{103}},
  \bibinfo{pages}{9401} (\bibinfo{year}{1995}).

\bibitem[{\citenamefont{Trudu et~al.}(2006)\citenamefont{Trudu, Donadio, and
  Parrinello}}]{TDP06}
\bibinfo{author}{\bibfnamefont{F.}~\bibnamefont{Trudu}},
  \bibinfo{author}{\bibfnamefont{D.}~\bibnamefont{Donadio}}, \bibnamefont{and}
  \bibinfo{author}{\bibfnamefont{M.}~\bibnamefont{Parrinello}},
  \bibinfo{journal}{Phys. Rev. Lett.} \textbf{\bibinfo{volume}{97}},
  \bibinfo{pages}{105701} (\bibinfo{year}{2006}).

\bibitem[{\citenamefont{Wang et~al.}(2007)\citenamefont{Wang, Gould, and
  Klein}}]{WGK07}
\bibinfo{author}{\bibfnamefont{H.}~\bibnamefont{Wang}},
  \bibinfo{author}{\bibfnamefont{H.}~\bibnamefont{Gould}}, \bibnamefont{and}
  \bibinfo{author}{\bibfnamefont{W.}~\bibnamefont{Klein}},
  \bibinfo{journal}{Phys. Rev. E} \textbf{\bibinfo{volume}{76}},
  \bibinfo{pages}{031604} (\bibinfo{year}{2007}).

\bibitem[{\citenamefont{Allen and Tildesley}(1987)}]{AT87}
\bibinfo{author}{\bibfnamefont{M.~P.} \bibnamefont{Allen}} \bibnamefont{and}
  \bibinfo{author}{\bibfnamefont{D.~J.} \bibnamefont{Tildesley}},
  \emph{\bibinfo{title}{Computer Simulation of Liquids}}
  (\bibinfo{publisher}{Clarendon}, \bibinfo{address}{Oxford},
  \bibinfo{year}{1987}).

\bibitem[{\citenamefont{Ibergay et~al.}(2007)\citenamefont{Ibergay, Ghoufi,
  Goujon, Ungerer, Boutin, Rousseau, and Malfreyt}}]{IGGUBRM07}
\bibinfo{author}{\bibfnamefont{C.}~\bibnamefont{Ibergay}},
  \bibinfo{author}{\bibfnamefont{A.}~\bibnamefont{Ghoufi}},
  \bibinfo{author}{\bibfnamefont{F.}~\bibnamefont{Goujon}},
  \bibinfo{author}{\bibfnamefont{P.}~\bibnamefont{Ungerer}},
  \bibinfo{author}{\bibfnamefont{A.}~\bibnamefont{Boutin}},
  \bibinfo{author}{\bibfnamefont{B.}~\bibnamefont{Rousseau}}, \bibnamefont{and}
  \bibinfo{author}{\bibfnamefont{P.}~\bibnamefont{Malfreyt}},
  \bibinfo{journal}{Phys. Rev. E} \textbf{\bibinfo{volume}{75}},
  \bibinfo{pages}{051602} (\bibinfo{year}{2007}).

\bibitem[{\citenamefont{Vrabec et~al.}(2006)\citenamefont{Vrabec, Kedia, Fuchs,
  and Hasse}}]{VKFH06}
\bibinfo{author}{\bibfnamefont{J.}~\bibnamefont{Vrabec}},
  \bibinfo{author}{\bibfnamefont{G.~K.} \bibnamefont{Kedia}},
  \bibinfo{author}{\bibfnamefont{G.}~\bibnamefont{Fuchs}}, \bibnamefont{and}
  \bibinfo{author}{\bibfnamefont{H.}~\bibnamefont{Hasse}},
  \bibinfo{journal}{Mol. Phys.} \textbf{\bibinfo{volume}{104}},
  \bibinfo{pages}{1509} (\bibinfo{year}{2006}).

\bibitem[{\citenamefont{Blander and Katz}(1972)}]{BK72}
\bibinfo{author}{\bibfnamefont{M.}~\bibnamefont{Blander}} \bibnamefont{and}
  \bibinfo{author}{\bibfnamefont{J.~L.} \bibnamefont{Katz}},
  \bibinfo{journal}{J. Stat. Phys.} \textbf{\bibinfo{volume}{4}},
  \bibinfo{pages}{55} (\bibinfo{year}{1972}).

\bibitem[{\citenamefont{Fenelonov et~al.}(2001)\citenamefont{Fenelonov,
  Kodenyov, and Kostrovsky}}]{FKK01}
\bibinfo{author}{\bibfnamefont{V.~B.} \bibnamefont{Fenelonov}},
  \bibinfo{author}{\bibfnamefont{G.~G.} \bibnamefont{Kodenyov}},
  \bibnamefont{and} \bibinfo{author}{\bibfnamefont{V.~G.}
  \bibnamefont{Kostrovsky}}, \bibinfo{journal}{J. Phys. Chem. B}
  \textbf{\bibinfo{volume}{105}}, \bibinfo{pages}{1050} (\bibinfo{year}{2001}).

\bibitem[{\citenamefont{Ford}(1997)}]{Ford97}
\bibinfo{author}{\bibfnamefont{I.~J.} \bibnamefont{Ford}},
  \bibinfo{journal}{Phys. Rev. E} \textbf{\bibinfo{volume}{56}},
  \bibinfo{pages}{5615} (\bibinfo{year}{1997}).

\bibitem[{\citenamefont{Schmelzer}(2001)}]{Schmelzer01}
\bibinfo{author}{\bibfnamefont{J.~W.~P.} \bibnamefont{Schmelzer}},
  \bibinfo{journal}{J. Coll. Interf. Sci.} \textbf{\bibinfo{volume}{242}},
  \bibinfo{pages}{354} (\bibinfo{year}{2001}).

\bibitem[{\citenamefont{Tolman}(1949)}]{Tolman49}
\bibinfo{author}{\bibfnamefont{R.~C.} \bibnamefont{Tolman}},
  \bibinfo{journal}{J. Chem. Phys.} \textbf{\bibinfo{volume}{17}},
  \bibinfo{pages}{333} (\bibinfo{year}{1949}).

\bibitem[{\citenamefont{Baidakov and Boltachev}(1999)}]{BB99}
\bibinfo{author}{\bibfnamefont{V.~G.} \bibnamefont{Baidakov}} \bibnamefont{and}
  \bibinfo{author}{\bibfnamefont{G.~S.} \bibnamefont{Boltachev}},
  \bibinfo{journal}{Phys. Rev. E} \textbf{\bibinfo{volume}{59}},
  \bibinfo{pages}{469} (\bibinfo{year}{1999}).

\bibitem[{\citenamefont{Laaksonen et~al.}(1994)\citenamefont{Laaksonen, Ford,
  and Kulmala}}]{LFK94}
\bibinfo{author}{\bibfnamefont{A.}~\bibnamefont{Laaksonen}},
  \bibinfo{author}{\bibfnamefont{I.~J.} \bibnamefont{Ford}}, \bibnamefont{and}
  \bibinfo{author}{\bibfnamefont{M.}~\bibnamefont{Kulmala}},
  \bibinfo{journal}{Phys. Rev. E} \textbf{\bibinfo{volume}{49}},
  \bibinfo{pages}{5517} (\bibinfo{year}{1994}).

\bibitem[{\citenamefont{Morris et~al.}(1994)\citenamefont{Morris, Wang, Ho, and
Chang}}]{MWHC94}
\bibinfo{author}{\bibfnamefont{J.~R.}~\bibnamefont{Morris}},
  \bibinfo{author}{\bibfnamefont{C.~Z.}~\bibnamefont{Wang}},
  \bibinfo{author}{\bibfnamefont{K.~M.}~\bibnamefont{Ho}}, \bibnamefont{and}
  \bibinfo{author}{\bibfnamefont{C.~T.}~\bibnamefont{Chan}},
  \bibinfo{journal}{Phys. Rev. B} \textbf{\bibinfo{volume}{49}},
  \bibinfo{pages}{3109} (\bibinfo{year}{1994}).

\bibitem[{\citenamefont{Reguera and Reiss}(2004)}]{RR04}
\bibinfo{author}{\bibfnamefont{D.} \bibnamefont{Reguera}}
  \bibnamefont{and} \bibinfo{author}{\bibfnamefont{H.}~\bibnamefont{Reiss}},
  \bibinfo{journal}{Phys. Rev. Lett.} \textbf{\bibinfo{volume}{93}},
  \bibinfo{pages}{165701} (\bibinfo{year}{2004}).

\bibitem[{\citenamefont{Lovett}(2007)}]{Lovett07}
\bibinfo{author}{\bibfnamefont{R.}~\bibnamefont{Lovett}},
  \bibinfo{journal}{Rep. Prog. Phys.} \textbf{\bibinfo{volume}{70}},
  \bibinfo{pages}{195} (\bibinfo{year}{2007}).

\bibitem[{\citenamefont{Schaaf et~al.}(2001)\citenamefont{Schaaf, Senger,
  Voegel, Bowles, and Reiss}}]{SSVBR01}
\bibinfo{author}{\bibfnamefont{P.}~\bibnamefont{Schaaf}},
  \bibinfo{author}{\bibfnamefont{B.}~\bibnamefont{Senger}},
  \bibinfo{author}{\bibfnamefont{J.-C.} \bibnamefont{Voegel}},
  \bibinfo{author}{\bibfnamefont{R.~K.} \bibnamefont{Bowles}},
  \bibnamefont{and} \bibinfo{author}{\bibfnamefont{H.}~\bibnamefont{Reiss}},
  \bibinfo{journal}{J. Chem. Phys.} \textbf{\bibinfo{volume}{114}},
  \bibinfo{pages}{8091} (\bibinfo{year}{2001}).

\bibitem[{\citenamefont{Salonen et~al.}(2007)\citenamefont{Salonen, Napari, and
  Vehkam{\"a}ki}}]{SNV07}
\bibinfo{author}{\bibfnamefont{M.}~\bibnamefont{Salonen}},
  \bibinfo{author}{\bibfnamefont{I.}~\bibnamefont{Napari}}, \bibnamefont{and}
  \bibinfo{author}{\bibfnamefont{H.}~\bibnamefont{Vehkam{\"a}ki}},
  \bibinfo{journal}{Mol. Sim.} \textbf{\bibinfo{volume}{33}},
  \bibinfo{pages}{245} (\bibinfo{year}{2007}).

\bibitem[{\citenamefont{Talanquer}(2007)}]{Talanquer07}
\bibinfo{author}{\bibfnamefont{V.}~\bibnamefont{Talanquer}},
  \bibinfo{journal}{J. Phys. Chem. B} \textbf{\bibinfo{volume}{111}},
  \bibinfo{pages}{3438} (\bibinfo{year}{2007}).

\bibitem[{\citenamefont{Rein ten Wolde and Frenkel}(1998)}]{WF98}
\bibinfo{author}{\bibfnamefont{P.}~\bibnamefont{Rein ten Wolde}}
  \bibnamefont{and} \bibinfo{author}{\bibfnamefont{D.}~\bibnamefont{Frenkel}},
  \bibinfo{journal}{J. Chem. Phys.} \textbf{\bibinfo{volume}{109}},
  \bibinfo{pages}{9901} (\bibinfo{year}{1998}).

\bibitem[{\citenamefont{Irving and Kirkwood}(1950)}]{IK50}
\bibinfo{author}{\bibfnamefont{J.~H.} \bibnamefont{Irving}} \bibnamefont{and}
  \bibinfo{author}{\bibfnamefont{J.~G.} \bibnamefont{Kirkwood}},
  \bibinfo{journal}{J. Chem. Phys.} \textbf{\bibinfo{volume}{18}},
  \bibinfo{pages}{817} (\bibinfo{year}{1950}).

\bibitem[{\citenamefont{Bernreuther and Vrabec}(2006)}]{BV05}
\bibinfo{author}{\bibfnamefont{M.}~\bibnamefont{Bernreuther}} \bibnamefont{and}
  \bibinfo{author}{\bibfnamefont{J.}~\bibnamefont{Vrabec}}, in
  \emph{\bibinfo{booktitle}{High Performance Computing on Vector Systems}},
  edited by \bibinfo{editor}{\bibfnamefont{M.}~\bibnamefont{Resch}}
  \bibnamefont{et~al.} (\bibinfo{publisher}{Springer}, \bibinfo{year}{2006}),
  pp. \bibinfo{pages}{187--195}.

\bibitem[{\citenamefont{Grottel et~al.}(2007)\citenamefont{Grottel, Reina,
  Vrabec, and Ertl}}]{GRVE07}
\bibinfo{author}{\bibfnamefont{S.}~\bibnamefont{Grottel}},
  \bibinfo{author}{\bibfnamefont{G.}~\bibnamefont{Reina}},
  \bibinfo{author}{\bibfnamefont{J.}~\bibnamefont{Vrabec}}, \bibnamefont{and}
  \bibinfo{author}{\bibfnamefont{T.}~\bibnamefont{Ertl}},
  \bibinfo{journal}{IEEE Trans. Vis. Comp. Graph.}
  \textbf{\bibinfo{volume}{13}}, \bibinfo{pages}{1624} (\bibinfo{year}{2007}).

\bibitem[{\citenamefont{Moroni et~al.}(2005)\citenamefont{Moroni, Rein ten Wolde,
and Bolhuis}}]{MWB05}
\bibinfo{author}{\bibfnamefont{D.}~\bibnamefont{Moroni}},
  \bibinfo{author}{\bibfnamefont{P.}~\bibnamefont{Rein ten Wolde}}, \bibnamefont{and}
  \bibinfo{author}{\bibfnamefont{P.~G.}~\bibnamefont{Bolhuis}},
  \bibinfo{journal}{Phys. Rev. Lett.}
  \textbf{\bibinfo{volume}{94}}, \bibinfo{pages}{235703} (\bibinfo{year}{2005}).

\bibitem[{\citenamefont{Sanz et~al.}(2007)\citenamefont{Sanz, Valeriani, Frenkel,
and Dijkstra}}]{SVFD07}
\bibinfo{author}{\bibfnamefont{E.}~\bibnamefont{Sanz}},
  \bibinfo{author}{\bibfnamefont{C.}~\bibnamefont{Valeriani}},
  \bibinfo{author}{\bibfnamefont{D.}~\bibnamefont{Frenkel}}, \bibnamefont{and}
  \bibinfo{author}{\bibfnamefont{M.}~\bibnamefont{Dijkstra}},
  \bibinfo{journal}{Phys. Rev. Lett.}
  \textbf{\bibinfo{volume}{99}}, \bibinfo{pages}{055501} (\bibinfo{year}{2007}).

\bibitem[{\citenamefont{Bhimalapuram et~al.}(2007)\citenamefont{Bhimalapuram,
  Chakrabarty, and Bagchi}}]{BCB07}
\bibinfo{author}{\bibfnamefont{P.}~\bibnamefont{Bhimalapuram}},
  \bibinfo{author}{\bibfnamefont{S.}~\bibnamefont{Chakrabarty}}, \bibnamefont{and}
  \bibinfo{author}{\bibfnamefont{B.}~\bibnamefont{Bagchi}},
  \bibinfo{journal}{Phys. Rev. Lett.} \textbf{\bibinfo{volume}{98}},
  \bibinfo{pages}{206104} (\bibinfo{year}{2007}).

\bibitem[{\citenamefont{Pierce et~al.}(1971)\citenamefont{Pierce, Sherman, and
  Mc{B}ride}}]{PSM71}
\bibinfo{author}{\bibfnamefont{T.}~\bibnamefont{Pierce}},
  \bibinfo{author}{\bibfnamefont{P.~M.} \bibnamefont{Sherman}},
  \bibnamefont{and} \bibinfo{author}{\bibfnamefont{D.~D.}
  \bibnamefont{Mc{B}ride}}, \bibinfo{journal}{Astronautica Acta}
  \textbf{\bibinfo{volume}{16}}, \bibinfo{pages}{1} (\bibinfo{year}{1971}).

\bibitem[{\citenamefont{Zahoranski et~al.}(1995)\citenamefont{Zahoranski,
  H{\"o}schele, and Steinwandel}}]{ZHS95}
\bibinfo{author}{\bibfnamefont{R.~A.} \bibnamefont{Zahoranski}},
  \bibinfo{author}{\bibfnamefont{J.}~\bibnamefont{H{\"o}schele}},
  \bibnamefont{and}
  \bibinfo{author}{\bibfnamefont{J.}~\bibnamefont{Steinwandel}},
  \bibinfo{journal}{J. Chem. Phys.} \textbf{\bibinfo{volume}{103}},
  \bibinfo{pages}{9038} (\bibinfo{year}{1995}).

\bibitem[{\citenamefont{Zahoranski et~al.}(1999)\citenamefont{Zahoranski,
  H{\"o}schele, and Steinwandel}}]{ZHS99}
\bibinfo{author}{\bibfnamefont{R.~A.} \bibnamefont{Zahoranski}},
  \bibinfo{author}{\bibfnamefont{J.}~\bibnamefont{H{\"o}schele}},
  \bibnamefont{and}
  \bibinfo{author}{\bibfnamefont{J.}~\bibnamefont{Steinwandel}},
  \bibinfo{journal}{J. Chem. Phys.} \textbf{\bibinfo{volume}{110}},
  \bibinfo{pages}{8842} (\bibinfo{year}{1999}).

\bibitem[{\citenamefont{Iland et~al.}(2007)\citenamefont{Iland, W{\"o}lk,
  Strey, and Kashchiev}}]{IWSK07}
\bibinfo{author}{\bibfnamefont{K.}~\bibnamefont{Iland}},
  \bibinfo{author}{\bibfnamefont{J.}~\bibnamefont{W{\"o}lk}},
  \bibinfo{author}{\bibfnamefont{R.}~\bibnamefont{Strey}}, \bibnamefont{and}
  \bibinfo{author}{\bibfnamefont{D.}~\bibnamefont{Kashchiev}},
  \bibinfo{journal}{J. Chem. Phys.} \textbf{\bibinfo{volume}{127}},
  \bibinfo{pages}{154506} (\bibinfo{year}{2007}).

\end{thebibliography}
\end{document}